\documentclass[10pt,letter,epsfig]{IEEEtran}
\usepackage{graphicx}
\usepackage{subfigure,color}
\usepackage{amssymb}
\usepackage{cite}
\usepackage{amsmath}
\usepackage{bm,comment}
\usepackage{algorithm}
\usepackage{algpseudocode}
\makeatletter

\newcommand{\Rmnum}[1]{\expandafter\@slowromancap\romannumeral #1@}
\newcommand{\mv}[1]{\mbox{\boldmath{$ #1 $}}}

\makeatother

\newtheorem{lemma}{Lemma}

\newtheorem{definition}{Definition}

\begin{document}
\title{Cooperative Beam Routing for Multi-IRS Aided Communication}
\author{Weidong Mei and Rui Zhang, \IEEEmembership{Fellow, IEEE}
\thanks{The authors are with the Department of Electrical and Computer Engineering, National University of Singapore, Singapore 117583 (e-mails: wmei@u.nus.edu, elezhang@nus.edu.sg). W. Mei is also with the NUS Graduate School for Integrative Sciences and Engineering, National University of Singapore, Singapore 119077.}}
\markboth{IEEE WIRELESS COMMUNICATIONS LETTERS}{}
\maketitle

\begin{abstract}
Intelligent reflecting surface (IRS) has been deemed as a transformative technology to achieve smart and reconfigurable environment for wireless communication. This letter studies a new IRS-aided communication system, where multiple IRSs assist in the communication between a multi-antenna base station (BS) and a remote single-antenna user by multi-hop signal reflection. Specifically, by exploiting the line-of-sight (LoS) link between nearby IRSs, a multi-hop cascaded LoS link between the BS and user is established where a set of IRSs are selected to successively reflect the BS's signal, so that the received signal power at the user is maximized. To tackle this new problem, we first present the closed-form solutions for the optimal active and cooperative passive beamforming at the BS and selected IRSs, respectively, for a given beam route. Then, we derive the end-to-end channel power, which unveils a fundamental trade-off in the optimal beam routing design between maximizing the multiplicative passive beamforming gain and minimizing the multi-reflection path loss. To reconcile this trade-off, we recast the IRS selection and beam routing problem as an equivalent shortest simple-path problem in graph theory and solve it optimally. Numerical results show significant performance gains of the proposed algorithm over benchmark schemes and also draw useful insights into the optimal beam routing design.
\end{abstract}

\section{Introduction}
As a new and promising application of digitally-controlled metasurface to wireless communications, intelligent reflecting surface (IRS) has received substantial attention recently. By dynamically tuning its massive reflecting elements, IRS realizes the fascinating concept of smart and reconfigurable environment for enhancing the wireless communication performance in assorted ways, such as passive relaying/beamforming, interference nulling/cancellation, channel statistics refinement, etc\cite{wu2019towards,basar2019wireless}. By efficiently integrating IRSs into wireless network, both its active and passive components will co-work in an intelligent way to bring a quantum-leap capacity improvement in anticipation. 

IRS has been extensively investigated for its performance optimization in various different wireless system setups (see, e.g., \cite{wu2020intelligent} and the references therein). However, most of the existing works on IRS considered the scenario where one or multiple distributed IRSs aid the communications between the base station (BS) and users via a single signal reflection or passive beamforming by one IRS only (e.g., \cite{huang2019reconfigurable,kammoun2020asymptotic,yu2020robust,hou2020reconfigurable,pan2020multicell,zhang2020intelligent,mei2020performance}). However, such a simplified approach may be ineffective in practice, e.g., when there are dense obstacles in the environment due to which one single IRS cannot help establish a line-of-sight (LoS) path between the BS and each user, regardless of the location of the IRS deployed. More importantly, the potential cooperative passive beamforming gain over the inter-IRS LoS link is not exploited, which results in suboptimal designs. To address the above issues, multiple IRSs can be employed to cooperatively serve each user via multi-hop signal reflection, so as to bypass scattered obstacles and exploit the inter-IRS cooperative beamforming gain. Motived by this, a double-IRS system was first proposed in \cite{han2020cooperative} where two IRSs, deployed near the BS and the user, respectively, cooperatively aid the transmission from a single-antenna BS to a single-antenna user via the inter-IRS double reflection link. It was revealed in \cite{han2020cooperative} that a cooperative passive beamforming gain increases with the total number of reflecting elements, denoted by $M$, in the order of ${\cal O}(M^4)$, which is higher than ${\cal O}(M^2)$ of the conventional single-IRS system with the single-reflection link only. This work was later extended to more practical/general scenarios of Rician fading channel and multi-antenna/multi-user systems in \cite{you2020wireless} and \cite{zheng2020double}, respectively. However, the general multi-IRS aided wireless communication with multi-hop (i.e., more than two hops) signal reflection/passive beamforming has not been studied in the literature yet, to the authors' best knowledge. 
\begin{figure}[!t]
\centering
\includegraphics[width=3.2in]{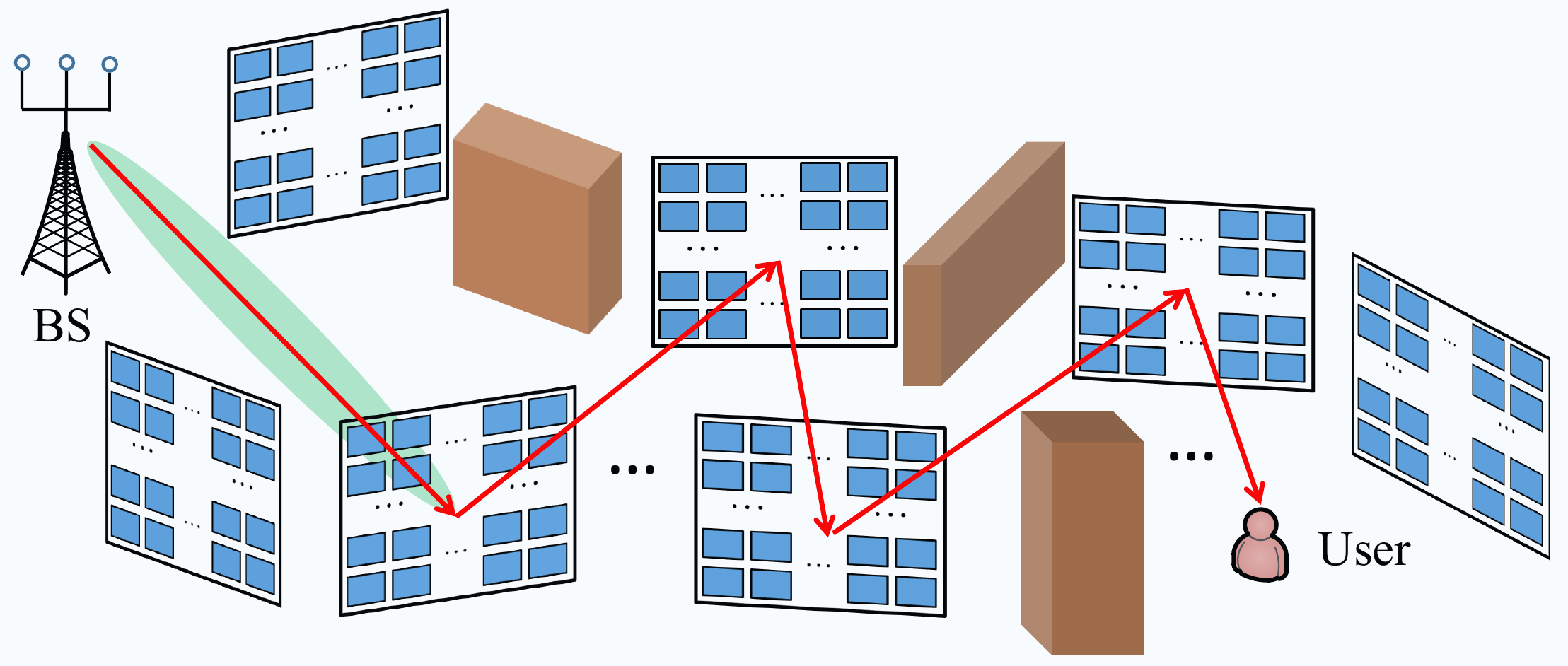}
\DeclareGraphicsExtensions.
\caption{A multi-IRS aided communication system with cooperative beam routing.}\label{MultiIRS}
\vspace{-15pt}
\end{figure}

To fill this gap, this letter considers a general multi-IRS aided wireless communication system as shown in Fig.\,\ref{MultiIRS}, where multiple IRSs assist in the downlink communication from a multi-antenna BS to a remote single-antenna user via successive signal reflection. By exploiting the short-distance LoS channel between nearby IRSs, an end-to-end equivalent LoS route can be established between the BS and user with a set of properly selected IRSs and their jointly designed passive beamforming along with the active beamforming at the BS, thus named as {\it cooperative beam routing}. It is worth noting that although this can be similarly achieved by the conventional multi-hop relaying with active relays, the proposed multi-IRS system is generally more energy and spectral efficient due to its passive, full-duplex and interference-free signal reflection\cite{wu2020intelligent}. Specifically, we aim to jointly optimize the selected IRSs or their reflection path with the BS/IRSs' active/passive beamforming to maximize the end-to-end equivalent channel power (see Fig.\,\ref{MultiIRS}). Note that selecting more IRSs helps reap a more pronounced multiplicative passive beamforming gain; while this may also incur higher path loss due to more signal reflections. Thus, there is an interesting trade-off to be reconciled in our proposed beam routing optimization. In this letter, we provide an optimal solution to this problem. First, we derive the optimal BS/IRSs' active/passive beamforming in closed-form to maximize the end-to-end channel power with a given reflection path by exploiting the inter-IRS LoS channels. Next, we recast the optimal path selection problem as an equivalent shortest simple-path problem (SSPP) in graph theory. Depending on whether negative weights exist in the constructed graph or not, we propose two customized algorithms to solve the equivalent SSPP problem optimally for both cases. Numerical results show that the proposed solution significantly outperforms other heuristic schemes.

{\it Notations:} For a complex number $s$, $s^*$, $\lvert s \rvert$ and $\angle s$ denote its conjugate, amplitude and phase, respectively. For a vector ${\mv a}$, ${\rm diag}({\mv a})$ denotes a diagonal matrix whose entries are the elements of $\mv a$, while $({\mv a})_m$ denotes its $m$-th entry. $a_{i,j}$ denotes the entry of a matrix $\mv A$ in the $i$-th row and the $j$-th column. $\lvert \Omega \rvert$ denotes the cardinality of a set $\Omega$. $n!$ denotes the factorial of a positive integer $n$. $\lfloor \cdot \rfloor$ denotes the greatest integer less than or equal to its argument.

\section{System Model and Beamforming Design}
\subsection{System Model}
As shown in Fig.\,\ref{MultiIRS}, we consider the downlink communication from a BS equipped with $N\,(>1)$ antennas to a single-antenna user in the presence of $J\,(>1)$ helping IRSs. We consider a challenging scenario where the direct BS-user link is severely blocked due to scattered obstacles. As such, the BS can only communicate with the user through a multi-reflection signal path that is formed by a set of selected IRSs via their cooperative passive beamforming over pairwise LoS channels. Assume that each IRS is equipped with $M$ passive reflecting elements. For convenience, we denote the sets of IRSs and reflecting elements per IRS as ${\cal J}\triangleq \{1,2,\cdots,J\}$ and ${\cal M}\triangleq \{1,2,\cdots,M\}$, respectively. We denote the reflection coefficient matrix of each IRS $j, j \in \cal J$ by ${\mv \Phi}_j={\rm diag}\{e^{j\theta_{j,1}},e^{j\theta_{j,2}},\cdots,e^{j\theta_{j,M}}\} \in {\mathbb C}^{M \times M}$, where we have set the reflection amplitude as one to maximize the reflected signal power by IRS and facilitate its hardware implementation\cite{huang2019reconfigurable,kammoun2020asymptotic,yu2020robust,hou2020reconfigurable,pan2020multicell,zhang2020intelligent,mei2020performance,han2020cooperative,you2020wireless,zheng2020double}. For convenience, we refer to the BS and user as nodes 0 and $J+1$ in the system, respectively. Accordingly, we define ${\mv H}_{0,j} \in {\mathbb C}^{M \times N}, j \in {\cal J}$ as the channel from the BS to IRS $j$, ${\mv g}_{j,J+1}^{H} \in {\mathbb C}^{1 \times M}, j \in {\cal J}$ as that from IRS $j$ to the user, and ${\mv S}_{i,j} \in {\mathbb C}^{M \times M}, i,j \in {\cal J}, i \ne j$ as that from IRS $i$ to IRS $j$.

Without loss of generality, we assume that the BS is equipped with a uniform linear array (ULA), while the passive reflecting elements of each IRS in $\cal J$ are arranged in a uniform rectangular array (URA) parallel to the $x$-$z$ plane. Denote by $M_1$ and $M_2$ the number of IRS reflecting elements in the vertical and horizontal directions of each IRS, respectively, with $M=M_1M_2$. Note that by carefully deploying the IRSs, LoS propagation can be achieved for some BS-IRS, inter-IRS, and IRS-user links if they are sufficiently close to each other. To describe the LoS availability between any two nodes in the considered system, we define an LoS condition matrix ${\mv L} \in \{0,1\}^{(J+2) \times (J+2)}$. In particular, the link between node $i$ and node $j$ consists of an LoS path if $l_{i,j}=1$; otherwise, $l_{i,j}=0$. Furthermore, we set its diagonal elements to zero, i.e., $l_{i,i}=0, \forall i$. Obviously, we have $l_{i,j}=l_{j,i}, \forall i,j$. This LoS condition matrix is assumed to be fixed and known after deploying the IRSs. As such, a multi-hop LoS channel can be established from the BS to the user by selecting a subset of IRSs from $\cal J$ in an order of reflection. For example, for the case of two selected IRSs, we can select IRSs $i$ and $j$ if $l_{0,i}=l_{i,j}=l_{j,J+1}=1$. The IRSs that are not selected can be switched off or regarded as random scatterers in the system, which may result in additional signal paths from the BS to the user. However, these randomly scattered paths generally have much lower strength as compared to the constructed LoS link thanks to the inter-IRS passive beamforming gain, especially for the case of practically large $M$\cite{mei2020performance}.

As such, we consider the LoS path only (if available) in each of ${\mv H}_{0,j}$'s, ${\mv g}_{j,J+1}$'s and ${\mv S}_{i,j}$'s, although other scattered multi-paths may also exist. Let $d_{i,j}, i \ne j$ denote the distance between nodes $i$ and $j$ (with some reference antenna/reflecting elements of the BS/IRSs selected without loss of generality). It is assumed that $d_{i,j} > 1$ meter (m), $\forall i \ne j$, to ensure the far-field propagation between nodes\cite{han2020cooperative}. The LoS channel between any two nodes is modeled as the product of array responses at two sides. For the ULA at the BS, the array response is expressed as ${\mv a}_B(\vartheta) \in {\mathbb R}^{N \times 1}$ with $({\mv a}_B(\vartheta))_n=e^{-j2\pi(n-1)d_A\sin\vartheta/\lambda}$, where $\vartheta$ denotes its angle-of-departure (AoD) relative to the BS antenna boresight, $d_A$ is the antenna spacing, and $\lambda$ is the carrier wavelength. While for the URA at each IRS, the array response is expressed as ${\mv a}_I(\vartheta^a,\vartheta^e) \in {\mathbb R}^{M \times 1}$ with $({\mv a}_I(\vartheta^a,\vartheta^e))_m=e^{-j2\pi d_I(\lfloor\! \frac{m-1}{M_1}\!\rfloor\sin\vartheta^e\cos\vartheta^a+(m-1-\lfloor\! \frac{m-1}{M_1}\!\rfloor M_1)\cos\vartheta^e)/\lambda}$, where $\vartheta^a$ and $\vartheta^e$ denote its azimuth angle-of-arrival (AoA)/AoD and elevation AoA/AoD, respectively, $d_I$ is the reflecting element spacing. Accordingly, we define $\vartheta_{0,j}$ as the AoD from the BS to IRS $j$, $\vartheta^a_{i,j}$/$\vartheta^e_{i,j}$ as the azimuth/elevation AoD from IRS $i$ to node $j$, and $\varphi^a_{j,i}$/$\varphi^e_{j,i}$ as the azimuth/elevation AoA at IRS $j$ from node $i$. The above AoAs and AoDs can be estimated by exploiting IRS controllers with transmitting/receiving function and/or equipping IRSs with dedicated sensors\cite{wu2020intelligent}.

Based on the above, we define ${\tilde{\mv h}}_{j,1}={\mv a}_B(\vartheta_{0,j})$ and ${\tilde{\mv h}}_{j,2}={\mv a}_I(\varphi^a_{j,0},\varphi^e_{j,0})$ for the LoS path/channel from the BS to IRS $j, j \in \cal J$, ${\tilde{\mv s}}_{i,j,1}={\mv a}_I(\vartheta^a_{i,j},\vartheta^e_{i,j})$ and ${\tilde{\mv s}}_{i,j,2}={\mv a}_I(\varphi^a_{j,i},\varphi^e_{j,i})$ for that from IRS $i$ to IRS $j, i,j \in \cal J$, as well as ${\tilde{\mv g}}_{j,J+1} = {\mv a}_I(\vartheta^a_{j,J+1},\vartheta^e_{j,J+1})$ for that from IRS $j$ to the user, $j \in {\cal J}$. Then, if $l_{0,j}=1$, the BS-IRS $j$ channel is given by
\begin{equation}\label{Ch1}
{\mv H}_{0,j} = \frac{\sqrt \beta}{d_{0,j}}e^{-\frac{j2\pi d_{0,j}}{\lambda}}{\tilde{\mv h}}_{j,2}{\tilde{\mv h}}^H_{j,1}, \;j \in {\cal J},
\end{equation}
where $\beta\, (<1)$ denotes the LoS path gain at the reference distance of 1 m. Similarly, if $l_{i,j}=1, i,j \in \cal J$, then the IRS $i$-IRS $j$ channel is expressed as
\begin{equation}\label{Ch2}
{\mv S}_{i,j} = \frac{\sqrt \beta}{d_{i,j}}e^{-\frac{j2\pi d_{i,j}}{\lambda}}{\tilde{\mv s}}_{i,j,2}{\tilde{\mv s}}^H_{i,j,1}, \;i, j \in {\cal J}, i \ne j.
\end{equation}
Finally, if $l_{j,J+1}=1$, then the IRS $j$-user channel is given by
\begin{equation}\label{Ch3}
{\mv g}^H_{j,J+1} = \frac{\sqrt \beta}{d_{j,J+1}}e^{-\frac{j2\pi d_{{j,J+1}}}{\lambda}}{\tilde{\mv g}}^H_{j,J+1}, \;j \in {\cal J}.
\end{equation}
Based on the above LoS channel characterization, we can construct a BS-user multi-reflection channel under given reflection path and BS/IRSs' active/passive beamforming, as shown next.\vspace{-6pt}

\subsection{Optimal Active and Passive Beamforming Design}
Let $\Omega=\{a_1,a_2,\cdots,a_K\}$ denote the reflection path from the BS to the user, where $K\, (\ge 1)$ and $a_k \in \cal J$ denote the number of selected IRSs/reflections and the index of the $k$-th selected IRS, $k=1,2,\cdots,K$, respectively. Then, $\Omega$ represents a feasible route if the following constraints are met:
\begingroup
\allowdisplaybreaks
\begin{align}
	&a_k \in {\cal J}, \;a_k \ne a_{k'}, \forall k,k' \in {\cal K}, k \ne k' \nonumber\\
	&l_{a_k,a_{k+1}}=1, \forall k \in {\cal K}, k \ne K \label{feasible}\\
	&l_{0,a_1}=l_{a_K,J+1}=1,\nonumber
\end{align}
where ${\cal K} \triangleq \{1,2,\cdots,K\}$. Note that the conditions in (\ref{feasible}) ensure that each IRS in $\cal J$ only reflects the BS's signal at most once and there is an LoS path between any two consecutive-reflection IRSs in $\Omega$ (besides the BS and IRS $a_1$ as well as IRS $a_K$ and the user). It is worth noting that the above wireless routing formed by active/passive beamforming resembles the celebrated circuit switching in the telephone network, for which the caller (BS) needs to establish a link with its callee (user) via multiple switches (IRSs) prior to communication.

Given a feasible route $\Omega$, we define $\mv w_B \in {\mathbb C}^{N \times 1}$ as the beamforming vector applied at the BS with $\lvert \mv w_B \rvert^2=1$. Then, the BS-user multi-reflection channel is expressed as
\begin{equation}\label{recvsig1}
h_{0,J+1}(\Omega)={\mv g}^H_{a_K,J+1}{\mv \Phi}_{a_K}\!\!\prod\limits_{k \in {\cal K}, k \ne K}\!\!\!\left({\mv S}_{a_k,a_{k+1}}{\mv \Phi}_{a_k}\right){\mv H}_{0,a_1}\mv w_B.
\end{equation}
It is noted that compared to the conventional system without IRS, the multi-reflection channel in (\ref{recvsig1}) depends on the cooperative passive beamforming design of the $K$ selected IRSs. By substituting (\ref{Ch1})-(\ref{Ch3}) into (\ref{recvsig1}) and rearranging the terms in it, we obtain
\begin{equation}\label{recvsig2}
h_{0,J+1}(\Omega)=e^{-\frac{j2\pi D(\Omega)}{\lambda}}\kappa(\Omega)\Bigg(\prod\limits_{k=1}^{K}A_k\Bigg){\tilde{\mv h}}^H_{a_1,1}\mv w_B,
\end{equation}
where
\begin{equation}\label{Ak}
A_k = \begin{cases}
	{\tilde{\mv s}}^H_{a_1,a_2,1}{\mv \Phi}_{a_1}{\tilde{\mv h}}_{a_1,2} &{\text{if}}\;\;k=1\\
	{\tilde{\mv s}}^H_{a_k,a_{k+1},1}{\mv \Phi}_{a_k}{\tilde{\mv s}}_{a_{k-1},a_k,2}&{\text{if}}\;\;2 \le k \le K-1\\
	\tilde{\mv g}^H_{a_K,J+1}{\mv \Phi}_{a_K}{\tilde{\mv s}}_{a_{K-1},a_K,2} &{\text{if}}\;\;k=K,
\end{cases}
\end{equation}
$D(\Omega)=d_{0,a_1}+d_{a_K,J+1}+\sum\nolimits_{k=1}^{K-1}d_{a_k,a_{k+1}}$ denotes the end-to-end transmission distance under the route $\Omega$, and
\begin{equation}\label{pathgain}
\kappa(\Omega)=\frac{(\sqrt\beta)^{K+1}}{d_{0,a_1}d_{a_K,J+1}\prod\limits_{k=1}^{K-1}d_{a_k,a_{k+1}}}
\end{equation}
denotes the cascaded LoS path gain between the BS and user under $\Omega$, which is the product of the LoS path gains of all constituent links under $\Omega$.

Thus, to maximize the BS-user equivalent channel power, $\lvert h_{0,J+1}(\Omega) \rvert^2$, the amplitude of each $A_k, k \in \cal K$ and ${\tilde{\mv h}}^H_{a_1,1}\mv w_B$ should be maximized by optimizing the phase shifts of IRS $a_k$, i.e., $\theta_{a_k,m}, m \in \cal M$, and the BS active beamforming $w_B$, respectively. According to (\ref{Ak}), to maximize each $\lvert A_k \rvert, k \in \cal K$, the phase shifts of each IRS $a_k$ should be set as\footnote{To focus on the cooperative beam routing design in this letter, we assume that the LoS channel between any two nodes, if it exists, is constant and known; while how to practically acquire such knowledge is an interesting problem to be addressed in future work.}
\begin{equation}\label{psall}
\theta_{a_k,m}\!=\!
\begin{cases}
	\angle(\tilde{\mv s}_{a_1,a_2,1})_m-\angle(\tilde{\mv h}_{a_1,2})_m &{\text{if}}\;k=1\\
	\angle(\tilde{\mv g}_{a_K,J+1})_m-\angle(\tilde{\mv s}_{a_{K-1},a_K,2})_m &{\text{if}}\;k=K\\
	\angle(\tilde{\mv s}_{a_k,a_{k+1},1})_m-\angle(\tilde{\mv s}_{a_{k-1},a_k,2})_m\!\!\! &{\text{otherwise}},
	\end{cases}
\end{equation}
for each $m \in \cal M$. This leads to $\lvert A_k \rvert = M, k \in \cal K$, as all array responses have unit-modulus entries. Moreover, to maximize $\lvert {\tilde{\mv h}}^H_{a_1,1}\mv w_B \rvert$, the optimal BS active beamforming is given by the maximum-ratio transmission (MRT) based on ${\tilde{\mv h}}_{a_1,1}$ in (\ref{recvsig2}), i.e.,
\begin{equation}\label{bmall}
	{\mv w}_B = e^{\frac{j2\pi D(\Omega)}{\lambda}}\frac{{\tilde{\mv h}}_{a_1,1}}{\lVert {\tilde{\mv h}}_{a_1,1} \rVert},
\end{equation}
and we have $\lvert {\tilde{\mv h}}^H_{a_1,1}\mv w_B \rvert = \sqrt{N}$.

With the phase shifts given in (\ref{psall}) and the active beamforming design given in (\ref{bmall}), it can be shown that the maximum BS-user equivalent channel power with any given feasible route $\Omega$ is expressed as
\begin{align}
\lvert h_{0,J+1}(\Omega) \rvert^2&=M^{2K}N\kappa^{2}(\Omega)\nonumber\\
&=\frac{M^{2K}N\beta^{K+1}}{d^2_{0,a_1}d^2_{a_K,J+1}\prod\limits_{k=1}^{K-1}d^2_{a_k,a_{k+1}}}.\label{eq1}
\end{align}

It is observed from (\ref{eq1}) that in addition to the conventional active BS beamforming gain of $N$, a multiplicative cooperative passive beamforming gain of $M^{2K}$ is achieved by $K$ selected IRSs, which increases exponentially with $K$. However, on the other hand, their multiple reflections also incur severe signal attenuation shown by $\beta^{K+1}$ in the numerator of (\ref{eq1}) as well as high ``product-distance'' power loss shown in the denominator of (\ref{eq1}), both of which increase with $K$ in general, thus resulting in lower end-to-end path gain $\kappa^{2}(\Omega)$ or higher end-to-end path loss $\kappa^{-2}(\Omega)$. It follows that the beam routing should be properly designed to balance the above trade-off, so as to maximize the end-to-end product channel power.\vspace{-6pt}

\section{Problem Formulation}\label{pf}
In this letter, we aim to maximize the end-to-end LoS channel power $\lvert h_{0,J+1}(\Omega) \rvert^2$ in (\ref{eq1}) by optimizing the reflection path $\Omega$, which specifies the selected IRSs and their order of reflection, subject to the feasibility constraints in (\ref{feasible}). The optimization problem is thus formulated as
\begin{equation}\label{op1}
{\text{(P1)}} \mathop {\max}\limits_{\{a_k\}_{k=1}^K,K}\; M^{2K}N\kappa^{2}(\Omega),\;\;\text{s.t.}\;\;{\text{(\ref{feasible})}}.
\end{equation}

However, (P1) is a non-convex combinatorial optimization problem due to its integer variables, which are coupled in the objective function in (\ref{op1}) and the constraints in (\ref{feasible}). Thus, it cannot be solved efficiently via standard optimization methods. In the following, we propose an optimal solution to it by resorting to the graph theory. \vspace{-3pt}

\section{Proposed Solution to (P1)}\label{sol}
In this section, we first recast (P1) as an equivalent SSPP in graph theory and then optimally solve it by leveraging two graph-optimization algorithms. \vspace{-6pt}

\subsection{Problem Reformulation}
Based on (\ref{eq1}), it is easy to see that maximizing $\lvert h_{0,J+1}(\Omega) \rvert^2$ in (P1) is equivalent to minimizing
\begin{equation}\label{eq2}
	\frac{1}{\lvert h_{0,J+1}(\Omega) \rvert^2}=\frac{M^2}{N}\cdot\frac{d^2_{0,a_1}}{M^2\beta}\cdot\frac{d^2_{a_K,J+1}}{M^2\beta}\prod\limits_{k=1}^{K-1}\frac{d^2_{a_k,a_{k+1}}}{M^2\beta}.
\end{equation}

Then, by taking the logarithm of (\ref{eq2}) and discarding irrelevant constant terms, (P1) becomes equivalent to
\begin{align}
\mathop {\min}\limits_{\{a_k\}_{k=1}^K,K}\;&\ln\frac{d_{0,a_1}}{M\sqrt\beta}+\ln\frac{d_{a_K,J+1}}{M\sqrt\beta}+\sum\limits_{k=1}^{K-1}\ln\frac{d_{a_k,a_{k+1}}}{M\sqrt\beta}\nonumber\\\quad\text{s.t.}\;\;&{\text{(\ref{feasible})}}.\label{op4}
\end{align}

Next, we recast problem (\ref{op4}) as an SSPP in graph theory subject to (\ref{feasible}). Specifically, we construct a directed weighted graph $G = (V,E)$. The vertex set $V$ is given by $V={\cal J} \cup \{0,J+1\}$. Furthermore, we assume that the beam can only be routed outwards from one IRS $i$ to a farther IRS $j$ from the BS, i.e., $d_{j,0} > d_{i,0}, i,j \in \cal J$, so as to reach the user as soon as possible. Accordingly, the edge set $E$ is defined as
\begin{align}
E=&\{(0,j)| l_{0,j} \!=\! 1, j \!\in\! {\cal J}\}\!\cup\!\{(i,j)|l_{i,j} \!=\! 1, d_{j,0} \!>\! d_{i,0}, i,j \!\in\! {\cal J}\} \nonumber\\
&\cup\{(j,J\!+\!1)|\, l_{j,J+1} = 1, j \in {\cal J}\},\label{edgeSet}
\end{align}
i.e., an edge exists from vertex $i$ to vertex $j$ if and only if there is an LoS path between them and $d_{j,0} > d_{i,0}$, except the case that vertex $j$ corresponds to the user, i.e., $j=J+1$. Furthermore, the weight of each edge $(i,j)$ in $E$ is set as $W_{i,j}=\ln\frac{d_{i,j}}{M\sqrt\beta}$. Note that $W_{i,j}$ may be negative if $d_{i,j} < M\sqrt{\beta}$. With graph $G$ constructed as above, we introduce the following definition.
\begin{definition}
	A simple path refers to the path in a graph which does not constitute repeated vertices.
\end{definition}

Based on Definition 1, a simple path from vertex $0$ to vertex $J+1$ in $G$ is able to satisfy all constraints in (\ref{feasible}). It follows from (\ref{edgeSet}) that $G$ is a directed acyclic graph (DAG). As such, any path in $G$ is a simple path. Then, problem (\ref{op4}) is equivalent to {\it finding the shortest path (i.e., with the minimum sum of the weights of the constituent edges) from vertex $0$ to vertex $J+1$ in $G$}. Next, we discuss the two cases without or with negative weights in $G$ to solve (P2) optimally in general.

\subsection{Optimal Solution to (P2)}\label{opsol}
{\bf{Case 1}}: The weights in $G$ are all non-negative, i.e., $W_{i,j}\ge 0, \forall (i,j) \in E$. In this case, (P2) can be optimally solved by applying the classical Dijkstra algorithm, which results in the worst-case complexity of ${\cal O}(\lvert E \rvert+\lvert V \rvert \log \lvert V \rvert)={\cal O}(\lvert E \rvert+ (J+2)\log (J+2))$ using the Fibonacci heap structure\cite{west1996introduction}.

{\bf{Case 2}}: There exists at least one negative weight in $G$. In this case, the Dijkstra algorithm may not return the shortest path as it is a greedy algorithm\cite{west1996introduction}. Nonetheless, the shortest path can be derived by applying some recursive algorithms, such as the Bellman-Ford algorithm\cite{west1996introduction} or solving the following {\it all hops shortest paths} (AHSP) problem\cite{cheng2004finding}. Specifically, instead of finding the shortest path directly with an unknown hop count, the AHSP problem aims to find, for all possible hop counts, the shortest paths from a given source to any other node in a graph. Specifically, denote by ${\cal N}_i = \{x | (x,i) \in E\}$ the set of all neighbors of node $i, i \in V$. Then, we have the following lemma.
\begin{lemma}[\!\!\cite{cheng2004finding}]\label{dp}
	Denote by $p^k(0,i)$ the $k$-hop shortest path from node 0 to another node $i \in V$. Then, if $k=0$, we have $p^k(0,i)=(0,i)$. If $(0,i)$ does not exist, we assume that there exists a virtual edge from node 0 to node $i$ with an infinite weight. For $k \ge 1$, $p^k(0,i)$ is the shortest path among the paths $p^{k-1}(0,x)+(x,i), x \in {\cal N}_i$, i.e., the paths resulted by concatenating $p^{k-1}(0,x)$ with the edge $(x,i)$ for all $x \in {\cal N}_i$. If there exists no $k$-hop path from node 0 to node $i$, we assume that there exists a virtual $k$-hop path with an infinite weight.
\end{lemma}

Lemma \ref{dp} indicates that $p^k(0,J+1)$ can be obtained by applying the dynamic programming (DP), based on the initial condition for $k=0$ and the recursion for $k \ge 1$. Since there are $J$ nodes between nodes 0 and $J+1$, the maximum possible hop count is $J$. As such, we need to find $p^k(0,J+1)$ for all $1 \le k \le J$ and compare their sums of edge weights to determine the globally shortest path. Hence, the worst-case complexity of the DP procedure is ${\cal O}(J\lvert E \rvert)$\cite{cheng2004finding}, which is comparable to that of the Bellman-Ford algorithm but higher than that of the Dijkstra algorithm.\vspace{-3pt}

\section{Numerical Results}\label{sim}
In this section, numerical results are provided to demonstrate the efficacy of the proposed beam routing design as compared with some benchmark routing schemes. Unless otherwise specified, the simulation settings are as follows. We consider an indoor multi-IRS system (such as a smart factory) with $J=10$. By following the corresponding LoS probability specified in \cite{3GPP38901}, we assume that LoS propagation can be achieved between nodes $i$ and $j$, i.e., $l_{i,j}=1, i,j \in V$, if its occurrence probability is greater than 0.9, or $d_{i,j} \le$ 12 m. Accordingly, the graph representation of the considered multi-IRS system, i.e., $G$, and the coordinates of all nodes (in (m)) are shown in Fig.\,\ref{route}(a). The weights $W_{i,j}, (i,j) \in E$ assuming $M=900$ are also shown in Fig.\,\ref{route}(a). It is observed that the weights of some edges are negative (as highlighted in red); thus, Case 2 in Section \ref{opsol} occurs under this setup. The number of BS antennas is set to $N=2$. Similar to \cite{han2020cooperative}, the carrier frequency is set as 5 GHz. Hence, the carrier wavelength is $\lambda = 0.06$ m and the LoS path gain at the reference distance of 1 m is $\beta=(\lambda/4\pi)^2 = -46$ dB. 

\begin{figure}[!t]
\centering
\includegraphics[width=3.4in]{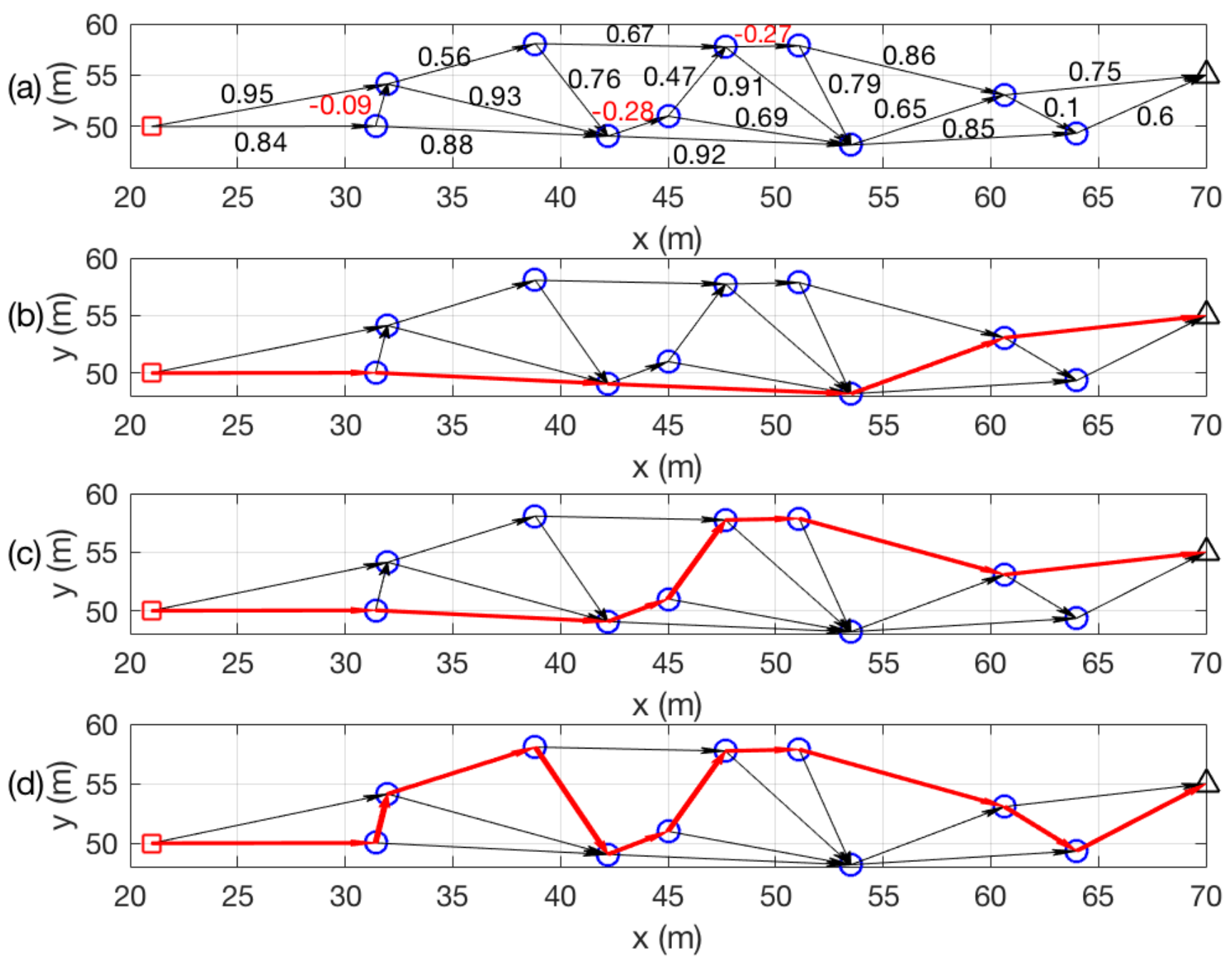}
\DeclareGraphicsExtensions.\vspace{-6pt}
\caption{(a) Graph representation of the simulation setup with $M=900$; (b) Optimal route with $M=400$; (c) Optimal route with $M=700$; (d) Optimal route with $M=1500$.}\label{route}
\vspace{-12pt}
\end{figure}
In Figs.\,\ref{route}(b)-\ref{route}(d), we plot the optimal beam routing solution for $M=400, 700$ and $1500$, respectively. It is observed that when $M=400$, the optimal route only goes through four IRSs before reaching the user. Moreover, it can be verified (via the Dijkstra algorithm) that this path also has the shortest distance among all paths from the BS to the user. This indicates that as $M=400$, maximizing the end-to-end path gain $\kappa^2(\Omega)$ (or minimizing the end-to-end path loss $\kappa^{-2}(\Omega)$) is dominant over maximizing the multiplicative passive beamforming gain $M^{2K}$ to maximize the effective channel power $\lvert h_{0,J+1}(\Omega) \rvert^2$ in (\ref{eq1}). However, as $M$ increases, it is observed that the optimal route makes a detour to select more IRSs, which is due to the more significant effect of the cooperative passive beamforming gain with increasing $M$. 

\begin{figure}[!t]
\centering
\includegraphics[width=3.5in]{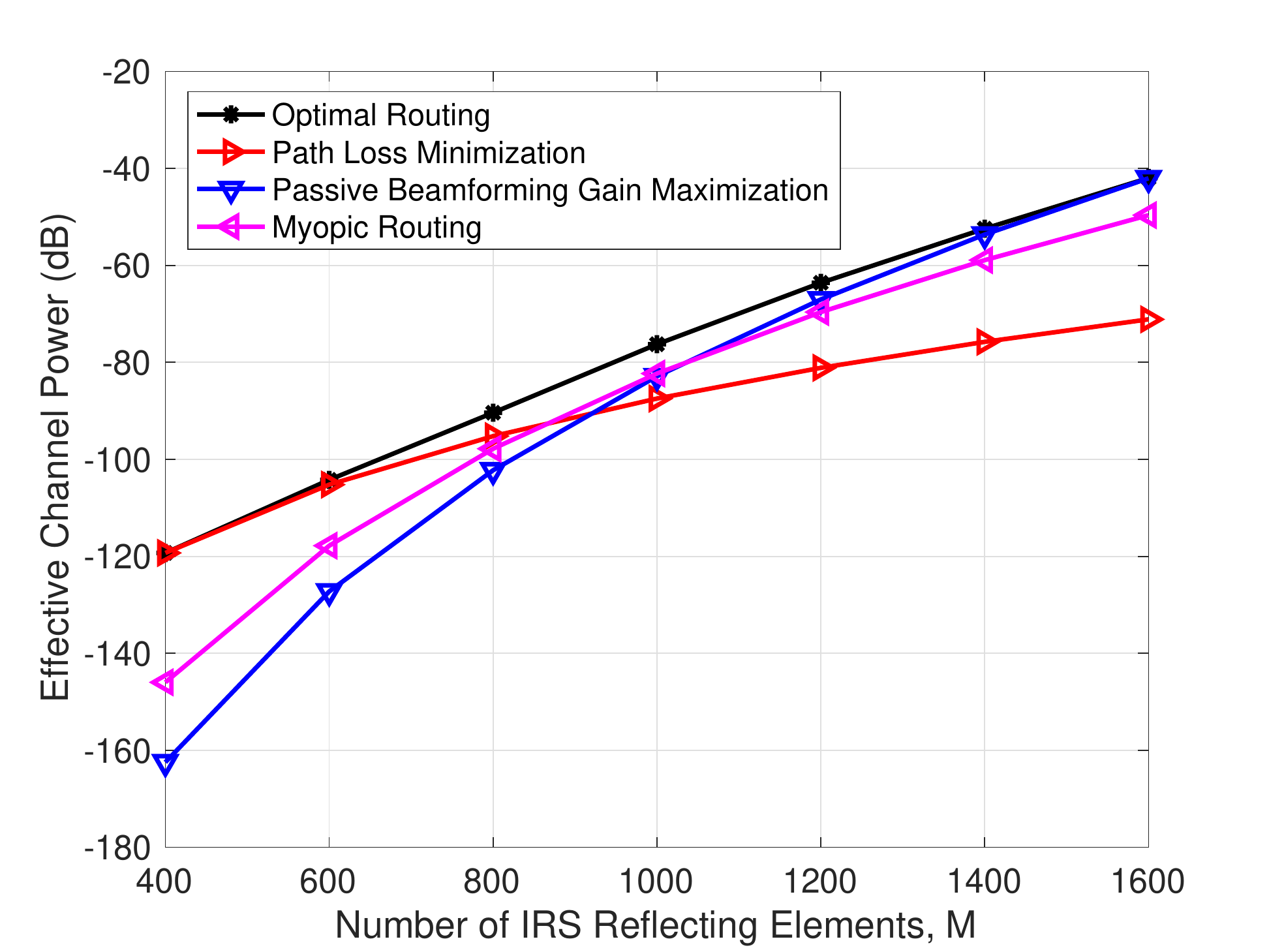}
\DeclareGraphicsExtensions.\vspace{-6pt}
\caption{Effective channel power versus number of IRS reflecting elements.}\label{ChPwvsM}
\vspace{-12pt}
\end{figure}
Next, we plot in Fig.\,\ref{ChPwvsM} the effective BS-user channel power, $\lvert h_{0,J+1}(\Omega) \rvert^2$, by different routing designs versus $M$. Specifically, we consider the following three benchmark schemes for comparison. The first and second benchmarks maximize the cooperative passive beamforming gain $M^{2K}$ (or the hop count $K$) and end-to-end path gain $\kappa^2(\Omega)$ in (\ref{eq1}), respectively, instead of balancing the trade-off between them as in the proposed routing design. The optimal route by the first benchmark can be derived as $p^{k'}(0,J+1)$ based on Lemma \ref{dp}, where $k'$ denotes the largest $k$ such that $p^k(0,J+1)$ exists. From Fig.\,\ref{route}(a), we obtain $k'=10$. The optimal route by the second benchmark can be easily obtained by assuming $M=1$ and then applying the proposed algorithms in Section \ref{opsol}. In addition, the third scheme is a myopic routing, whereby the next IRS is always selected as the closest one to the current IRS (or the BS as the starting vertex) among all unselected IRSs, until the user is reached. It is observed from Fig.\,\ref{ChPwvsM} that the proposed optimal routing solution significantly outperforms the myopic benchmark over the whole range of $M$ considered. While the first and second benchmarks are observed to achieve close performance to the proposed solution when $M$ is large and small, respectively, due to different dominating effects of cooperative passive beamforming gain and end-to-end path loss. However, when $M$ is moderate, it is observed that these two benchmarks result in much worse performance than the proposed routing solution, which optimally balances the trade-off between maximizing the cooperative passive beamforming gain and minimizing the end-to-end path loss. \vspace{-4pt}

\section{Conclusions}
This letters studies a multi-IRS aided communication system, where a cascaded LoS link is established between a BS and a remote user by leveraging the successive signal reflections of multiple selected IRSs. We present the optimal BS active and IRSs' passive beamforming design as well as the optimal IRS selection and beam routing solution, which maximize the cascaded LoS channel power by leveraging graph theory. It is shown that there exists a fundamental trade-off in the optimal routing design between minimizing the end-to-end path loss and maximizing the cooperative passive beamforming gain, where the former/latter has a more dominating effect when the number of IRS reflecting elements is small/large.\vspace{-6pt}

\bibliography{IRScoop}
\bibliographystyle{IEEEtran}
\end{document}